\documentclass[aps,pra,twocolumn,amsmath,amssymb,showpacs,nofootinbib]{revtex4}

\newcommand{\ket}[1]{|{#1}\rangle}
\newcommand{\bra}[1]{\langle{#1}|}
\usepackage[dvips]{graphicx}
\usepackage{mathrsfs}

\begin{document}

\title{Non-deterministic approximation of photon number discriminating detectors using non-discriminating detectors}

\author{Peter P. Rohde}
\email[]{rohde@physics.uq.edu.au}
\homepage{http://www.physics.uq.edu.au/people/rohde/}
\affiliation{Centre for Quantum Computer Technology, Department of Physics\\ University of Queensland, Brisbane, QLD 4072, Australia}

\date{\today}

\begin{abstract}
We present a scheme for non-deterministically approximating photon number resolving detectors using non-discriminating detectors. The model is simple in construction and employs very few physical resources. Despite its non-determinism, the proposal may nonetheless be suitable for use in some quantum optics experiments in which non-determinism can be tolerated. We analyze the detection scheme in the context of an optical implementation of the controlled-NOT gate, an inherently non-deterministic device. This allows the gate's success probability to be traded away for improved gate fidelity, assuming high efficiency detectors. The scheme is compared to two other proposals, both deterministic, for approximating discriminating detectors using non-discriminating detectors: the cascade and time division multiplexing schemes.
\end{abstract}

\pacs{03.67.Lx,42.50.-p}

\maketitle

\section{INTRODUCTION}
Photon number resolving photo-detection is a necessary prerequisite for many interesting quantum optics experiments. In particular, linear optics quantum computing (LOQC) \cite{bib:KLM} requires discriminating photo-detection, \mbox{\emph{i.e.}} the ability to discriminate not just between the presence or absence of photons, but the actual number of photons. Unfortunately such detectors are not presently available. To remedy this, several schemes for approximating photon number resolving detectors using non-discriminating detectors have been proposed, including cascade networks \cite{bib:Kok,bib:Paul,bib:Bartlett} and time division multiplexing (TDM) \cite{bib:Achilles,bib:Banaszek}. In this paper we propose an extremely simple scheme for non-deterministically approximating photon number resolving detectors using only a small number of low reflectivity beamsplitters and non-discriminating photo-detectors. The proposal differs from previous proposals in that fewer physical resources are required (\mbox{\emph{i.e.}} a small number of beamsplitters and photo-detectors) and the scheme is not very prone to photon loss. The disadvantage is that it only succeeds with low probability, making its applications very limited. However, despite this drawback the scheme could be useful for LOQC and other experiments in which non-determinism is not necessarily problematic. As with all other such schemes, the proposal requires very high efficiency photo-detectors to work reliably.

We begin by reviewing the cascade and TDM photo-detection schemes. We then introduce our proposal for non-deterministically approximating photon number resolving detection using non-discriminating detectors. Finally we present simulated results illustrating the effectiveness of the scheme in the context of the simplified Knill-Laflamme-Milburn (KLM) controlled-NOT (CNOT) gate proposed by Ralph \mbox{\emph{et. al.}} \cite{bib:Ralph}, a fundamental 2-qubit LOQC gate. In this context gate fidelity can be made arbitrarily high, at the expense of success probability, provided that high efficiency non-discriminating photo-detectors are available.

\section{CASCADE PHOTO-DETECTION}
\begin{figure}[!htb]
\includegraphics[width=0.2\textwidth]{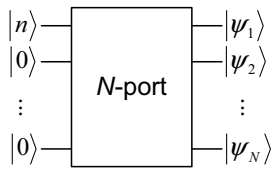}
\caption{\label{fig:N_port}An $N$-port device which takes an input Fock state and produces $N$ output states, $\ket{\psi_1}\dots\ket{\psi_N}$. The arrival statistics of incident photons are uniform across the outputs.}
\end{figure}

We begin by reviewing the cascade photo-detection scheme \cite{bib:Kok,bib:Paul,bib:Bartlett}. The basic idea behind this proposal is to use a network of beamsplitters, commonly referred to as an $N$-port, shown in Figure \ref{fig:N_port}, to divide a single input mode into many output modes, each of which is followed by a non-discriminating photo-detector. The device is constructed such that if a single photon enters the device, it has an equal probability of triggering any one of the photo-detectors. If multiple photons are incident on the device and there are a large number of output ports, there is a very low probability that two or more photons will reach the same detector. Thus the number of photons in an input Fock state can be approximated as the sum of how many detectors \emph{clicked}. Clearly this scheme requires high efficiency detectors to work effectively, since undetected photons will skew the perceived photon count. In fact this is a requirement for all of the number resolving detection schemes discussed in this paper. Another potential problem which can upset the perceived number of photons is that of \emph{dark counts}, whereby a detector \emph{clicks} even if no photon is incident upon it. In all of our discussions we neglect the effects of dark counts.

\begin{figure}[!htb]
\includegraphics[width=0.25\textwidth]{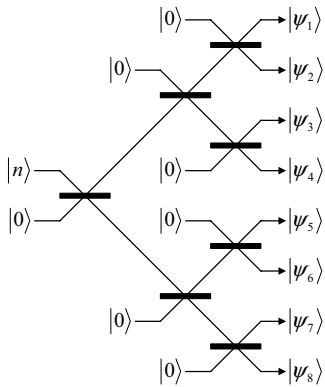}
\caption{\label{fig:tree_network}Construction of an $N$-port, for $N=8$, using a tree network of $\eta=0.5$ beamsplitters.}
\end{figure}

There are numerous ways in which an $N$-port device could be constructed from beamsplitters. The most obvious construction is to employ a tree network of 50/50 beamsplitters as shown in Figure \ref{fig:tree_network}. This requires $N-1$ beamsplitters, where $N$ is the number of output ports and is an integer power of 2. The actual internal construction of the $N$-port however is irrelevant. All that matters is the arrival statistics at the output ports.

It is evident that in order to achieve high confidence that the number of perceived photons was the actual number of photons present in the incident Fock state, not only are high efficiency detectors required, but also a large number of output ports in order to attain sufficient suppression of higher photon number terms at the detectors. Hence the cascade network has the drawback that a relatively large number of both beamsplitters and photo-detectors are necessary, which can be experimentally prohibitive.

An expression can be derived \cite{bib:Kok} for the probability that the measured number of photons is equal to the actual number of incident photons. We use $n$ to denote the number of incident photons (\mbox{\emph{i.e.}} $\ket{\psi}_{in}=\ket{n}$) and $m$ the measured or perceived number of photons. Allowing for detector inefficiency\footnote{Detector inefficiency is modeled by introducing beamsplitters prior to the detectors with reflectivity equal to their quantum efficiency. The \emph{loss modes} are not observed and thus are traced out.} the expression is
\begin{equation}
P(m=n)=\frac{{\eta_{eff}}^{n}N!}{N^n(N-n)!}
\end{equation}
where $\eta_{eff}$ is the quantum efficiency of the photo-detectors. It is evident that in the limit of high detector efficiency and a large number of beamsplitters, the probability that the measurement result is correct approaches unity, \mbox{\emph{i.e.}}
\begin{equation}
\lim_{\eta_{eff}\to1}P(m=n)=1\ \ \mathrm{for}\ \ N\gg n
\end{equation}

\section{TIME DIVISION MULTIPLEXING PHOTO-DETECTION}
\begin{figure}[!htb]
\includegraphics[width=0.4\textwidth]{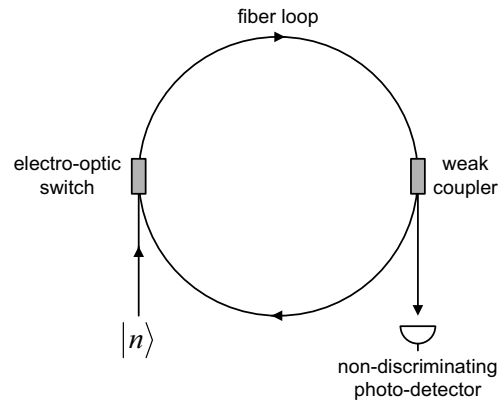}
\caption{\label{fig:TDM}Time division multiplexing photo-detection using a fiber loop, a weak coupler, an electro-optic switch and a non-discriminating photo-detector.}
\end{figure}

We now review the time division multiplexing photo-detection scheme \cite{bib:Achilles,bib:Banaszek}. This is a logical extension of the cascade approach which, instead of employing beamsplitters, utilizes a low loss fiber loop which inefficiently couples out to a non-discriminating photo-detector, shown in Figure \ref{fig:TDM}. The principle is similar to the cascade network. The Fock state loops around indefinitely inside the fiber loop and after each round-trip there is some probability, determined by the strength of the coupler, that a given photon will couple out of the loop and into the detector. Thus the device divides a Fock state into \emph{time-bins} which are measured independently of one anther. This is in contrast to the cascade approach whereby the input state is divided into \emph{spatial-bins}. The advantage of this scheme is that the probability of two or more photons triggering a single detector event can easily be reduced simply by lowering the strength of the coupler, without any physical resource overhead. In the cascade scheme this would require the addition of extra beamsplitters and photo-detectors. The disadvantage is that all fibers are inherently lossy to some degree which exacerbates the problem that photon loss may corrupt the perceived photon number. As with the cascade scheme, the TDM approach requires high efficiency photo-detectors to be accurate.

\section{NON-DETERMINISTIC PHOTO-DETECTION}
We now introduce our proposal for non-deterministically approximating photon number resolving detectors. In fact the proposal is a class of detectors, which, when they \emph{click}, indicate that with high probability exactly the desired number of photons were incident on the device.

We begin by explaining the simplest case: a detector which sometimes \emph{clicks} when $n=1$, but with extremely low probability when $n\neq1$. In other words, a measurement of $m=1$ tells us with a high level of confidence that $n=1$.

\begin{figure}[!htb]
\includegraphics[width=0.4\textwidth]{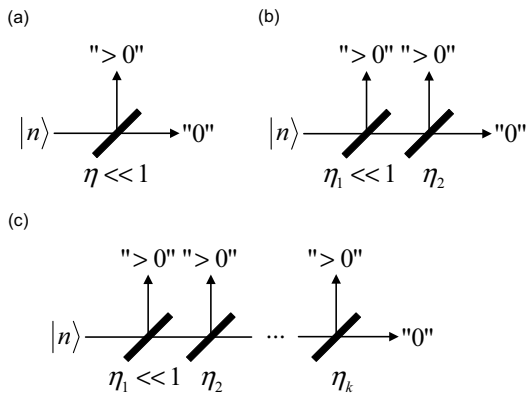}
\caption{\label{fig:NDPDM}Non-deterministic photo-detection scheme for the 1-photon (a), 2-photon (b) and $k$-photon (c) cases. ``0'' indicates conditioning upon no detector \emph{click} and ``$>$0'' upon a detector \emph{click} (\mbox{\emph{i.e.}} detecting at least one photon).}
\end{figure}

The scheme is illustrated in Figure \ref{fig:NDPDM}(a), which operates as follows. The incident Fock state is passed through a very low reflectivity beamsplitter and a photo-detector placed at each beamsplitter output. We condition upon detecting no photons in the transmitted mode. When the photo-detector associated with the reflected mode \emph{clicks} we can say with high probability that there was exactly one photon present in the incident state, since the probability that multiple photons are reflected from the low reflectivity beamsplitter is very low. Clearly the device also has very low success probability, since most of the time incident photons will be transmitted through the beamsplitter, thereby triggering the ``0'' detector.

We derive an expression for the probability of perceiving one photon for an $n$-photon input state as follows. We let $P_{trig}$ denote the probability that a single photon triggers the ``$>$0'' detector and $P_{loss}$ the probability that a given photon is lost, which arises due to detector inefficiencies. The probability that we measure exactly one photon is simply the sum over all possible outcomes whereby at least one photon triggers the ``$>$0'' detector and no photons trigger the ``0'' detector. This can be expressed as a binomial sum in terms of the probabilities $P_{trig}$ and $P_{loss}$ as
\begin{equation}
P(m=1)=\sum_{i=1}^n\binom{n}{i}{P_{trig}}^i{P_{loss}}^{n-i}
\end{equation}
We recognize that \mbox{$P_{trig}=\eta_{ref}\eta_{eff}$} and \mbox{$P_{loss}=1-\eta_{eff}$}, where $\eta_{ref}$ is the reflectivity of the beamsplitter and $\eta_{eff}$ is again the quantum efficiency of the detectors. Hence the expression for the probability of perceiving one photon is 
\begin{eqnarray}
P(m=1)&=&\sum_{i=1}^n\binom{n}{i}{\eta_{ref}}^i{\eta_{eff}}^i(1-\eta_{eff})^{n-i}\nonumber\\
&=&(1-\eta_{eff})^n\Big[\Big(1+\frac{\eta_{eff}\eta_{ref}}{1-\eta_{eff}}\Big)^n-1\Big]
\end{eqnarray}

The scheme can be logically generalized to construct $k$-photon discriminating detectors. We simply introduce a linear chain of low reflectivity beamsplitters and condition upon detecting no photon at the mode transmitted at the end of the chain and at least one photon at each of the reflected modes. Figure \ref{fig:NDPDM} illustrates this generalization for the 2-photon (b) and $k$-photon (c) cases. For optimality we require that photon arrival statistics be uniform across the ``$>$0'' modes. A sufficient condition for ensuring this is
\begin{equation}
\eta_i=\frac{\eta_{i-1}}{1-\eta_{i-1}}
\end{equation}
where $\eta_i$ is the reflectivity of the $i^\mathrm{th}$ beamsplitter. The expression for the probability of detecting $k$ photons, $P(m=k)$, generalizes to a multinomial distribution,
\begin{widetext}
\begin{eqnarray}
P(m=k)&=&\sum_{n_1\geq1,n_2\geq1,\dots,n_k\geq1}(n_1,\dots,n_k,n_{loss})!{P_{trig}}^{n_{trig}}{P_{loss}}^{n_{loss}}\nonumber\\
&=&\sum_{n_1\geq1,n_2\geq1,\dots,n_k\geq1}(n_1,\dots,n_k,n_{loss})!(\eta_{ref}\eta_{eff})^{n_{trig}}(1-\eta_{eff})^{n_{loss}}
\end{eqnarray}
\end{widetext}
where $n_i$ is the number of photons triggering the $i^\mathrm{th}$ ``$>$0'' detector, \mbox{$n_{trig}=\sum_{i=1}^{k}n_i$} is the number of photons which trigger a detector, \mbox{$n_{loss}=n-\sum_{i=1}^{k}n_i$} is the number of photons lost due to detector inefficiency, and we are summing over all possible combinations of $n_i$ such that every ``$>$0'' detector \emph{clicks} and the ``0'' detector does not.

\begin{figure}[!htb]
\includegraphics[width=0.5\textwidth]{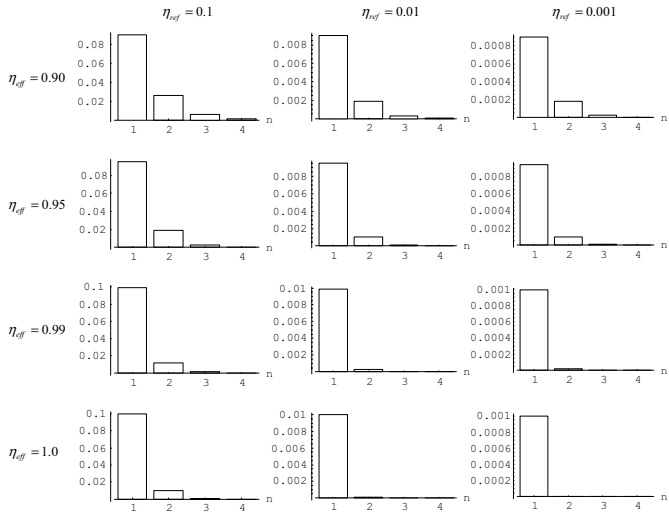}
\caption{\label{fig:suppression}Probability of perceiving one photon, \mbox{$P(m=1)$}, given an $n$-photon input state, against detector efficiency (rows) and beamsplitter reflectivity (columns). Note that the plots have different scales.}
\end{figure}

Figure \ref{fig:suppression} shows plots of the probability of perceiving one photon, $P(m=1)$, against the number of photons in the incident Fock state, for different detector efficiency and beamsplitter reflectivity scenarios. It is immediately obvious that the ability of the device to suppress higher photon number terms ($n\geq2$) is highly dependent upon both detector efficiency and beamsplitter reflectivity. In order to achieve very high confidence that $m=1$ implies $n=1$ we require detectors with very high efficiency (well above 90\%), beyond what is possible using presently available detectors, and a beamsplitter with very low reflectivity, which directly trades away the success probability of the device. However, within these constraints, the scheme is capable of very high suppression of higher order photon terms. For $k\geq2$ detectors we observe similar plots, albeit with an even greater dependence upon detector efficiency and beamsplitter reflectivity. This leads us to the conclusion that the proposed scheme would only be suitable for small $k$ detectors.

\section{APPLICATION TO THE NON-DETERMINISTIC HERALDED CNOT GATE}
We consider the non-deterministic photo-detection scheme in the context of the simplified KLM CNOT gate \cite{bib:KLM,bib:Ralph}. The gate employs the dual-rail convention, whereby each logical qubit is encoded across two orthogonal modes\footnote{In the logical basis this corresponds to \mbox{$\ket{0}_L\equiv\ket{0}_H\ket{1}_V$} and \mbox{$\ket{1}_L\equiv\ket{1}_H\ket{0}_V$}, where $H$ and $V$ label the two modes.}.

\begin{figure}[!htb]
\includegraphics[width=0.5\textwidth]{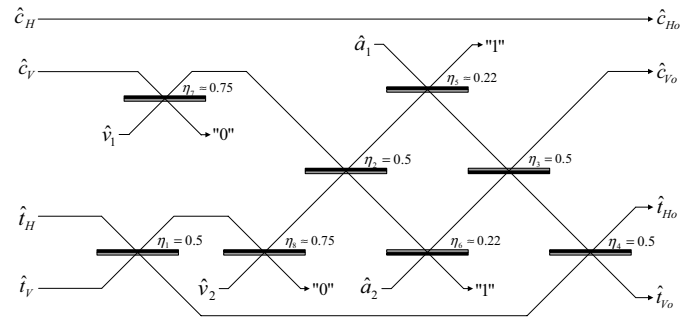}
\caption{\label{fig:CNOT_gate}Schematic of the simplified KLM CNOT gate using beamsplitters with reflectivities as indicated. ``0'' indicates conditioning upon detecting no photons, and ``1'' upon exactly one photon. $\hat{v}_1$ and $\hat{v}_2$ are vacuum inputs, while $\hat{a}_1$ and $\hat{a}_2$ are single photon ancilla inputs.}
\end{figure}

The gate is constructed from beamsplitters whose layout is illustrated in Figure \ref{fig:CNOT_gate}. The success of the gate is conditional upon detection of exactly one photon at each of the outputs labeled ``1'' and no photons at the outputs labeled ``0'', a procedure which succeeds approximately 5\% of the time.
Despite being non-deterministic, the gate can be used for efficient, scalable quantum computation, with the introduction of a quantum teleportation and quantum error correction procedure \cite{bib:KLM,bib:Gottesman}, which effectively circumvents the gate's non-determinism.

\begin{figure}[!htb]
\includegraphics[width=0.5\textwidth]{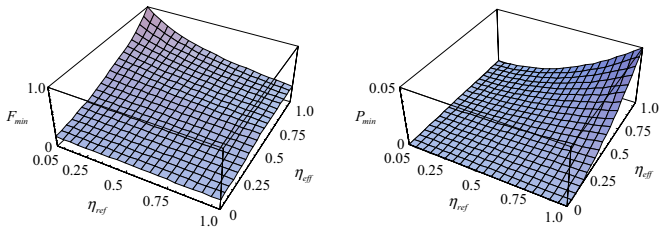}
\caption{\label{fig:CNOT_results}Worst-case fidelity and success probability of the CNOT gate against detector efficiency and beamsplitter reflectivity in the measurement circuit.}
\end{figure}

We apply our $m=1$ non-deterministic photo-detector model in place of the ``1'' conditional measurements and calculate the worst-case gate fidelity and success probability, defined as
\begin{eqnarray}
F_{min}&=&\frac{\bra{\psi}\hat{\rho}\ket{\psi}}{\mathrm{tr}(\hat{\rho})}\nonumber\\
P_{min}&=&\mathrm{tr}({\hat{\rho}})
\end{eqnarray}
where $\ket{\psi}$ is the ideal-case output state, defined according to the CNOT logical transformation, $\hat{\rho}$ is the density operator of the actual output state, and we minimize across all possible input states. Figure \ref{fig:CNOT_results} illustrates these measures, plotted against the quantum efficiency of the photo-detectors and the reflectivity of the beamsplitters in the measurement circuits.

In the limit as \mbox{$\eta_{eff}\to1$} and \mbox{$\eta_{ref}\to0$}, $F_{min}$ approaches unity. However, this comes at the expense of success probability, which, in the same limits, approaches zero. Gate fidelity is highly dependent upon detector efficiency. Unfortunately this dependence is sufficiently strong that acceptable gate fidelity would not be possible using present day photo-detectors. As an example, using 99\% efficient detectors we could achieve a worst-case gate fidelity of 95\% using a $\eta_{ref}=0.011$ beamsplitter, which would have a worst-case success probability of 0.0006\%. With 99.9\% efficient detectors we could achieve the same fidelity using a $\eta_{ref}=0.022$ beamsplitter, with a worst-case success probability of 0.0025\%. Clearly such high efficiency detectors are far beyond what is presently possible. Nonetheless, with the future in mind, high efficiency non-discriminating detectors is still far less demanding than high efficiency number resolving detectors. Note that implementing the CNOT gate using non-discriminating detectors corresponds to taking an \mbox{$\eta_{ref}=1.0$} cross-section through the plots in Figure \ref{fig:CNOT_results}.

\section{CONCLUSION}
We presented a model for constructing non-deterministic photon number resolving detectors using only a small number of non-discriminating detectors and low reflectivity beamsplitters. Compared to other schemes for approximating number resolving detectors, the physical resources required are modest. The proposal is also more resilient against photon loss than other schemes. The inherent non-determinism of the proposal limits its applicability enormously. However, despite this drawback, it could find uses in many interesting quantum optics experiments, including LOQC ones, in which non-determinism can be tolerated. In the context of the simplified KLM CNOT gate, we demonstrated that the proposed scheme could allow one to trade-off gate success probability for fidelity to an arbitrary degree, assuming high efficiency non-discriminating detectors are available. As with all such schemes, the proposal is highly dependent upon photo-detector efficiency. This dependence increases rapidly with the number of photons one attempts to discriminate, leading to the conclusion that the proposal would most likely only ever be practical for one or two photon discriminating detection.

\begin{acknowledgments}
We acknowledge Timothy Ralph and Geoffrey Pryde for helpful discussions. This work was supported by the Australian Research Council.
\end{acknowledgments}

\bibliography{paper.bib}

\end{document}